\documentclass[12pt]{article}
\pdfoutput=1
\usepackage{epsfig}
\usepackage{amsmath}
\usepackage{color}
\usepackage{booktabs}
\usepackage{amssymb}
\usepackage{subfigure}
\usepackage{comment}

\usepackage{setspace}
\usepackage{array}
\usepackage{times}
\usepackage{bm}
\usepackage{graphicx}
\usepackage{latexsym}

\newtheorem{theorem}{Theorem}[section]
\newtheorem{corollary}{Corollary}[section]
\newtheorem{proposition}{Proposition}[section]

\newtheorem{lemma}{Lemma}[section]

\newtheorem{remark}{Remark}[section]
\newtheorem{definition}{Definition}[section]
\newtheorem{assumption}{Assumption}[section]
\newtheorem{exmp}{Example}[section]

\newcommand{\ind}[1]{I{\left\{ #1 \right\}}}
\newcommand{\cond}{\;\middle\vert\;}
\newcommand{\expv}[1] {E\left( #1 \right)}
\newcommand{\pr}[1] {\mathrm{Pr}\left( #1 \right)}

\newcommand{\maxR}{(R \vee 1)}
\def\qed{\rule{2mm}{2mm}}

\usepackage[dvips, letterpaper, nohead, top=1.3in, bottom=1.3in, left=1.3in, right=1.3in]{geometry}

\begin{document}

\small\normalsize

\title{A New Approach for Large Scale Multiple Testing \\ with Application to FDR Control for Graphically \\ Structured Hypotheses\thanks{The authors are listed in alphabetical order.}}

\author{
 Wenge Guo\thanks{The research of the author was supported in part
by NSF Grant DMS-1309162.} \\
Department of Mathematical Sciences\\
New Jersey Institute of Technology \\
Newark, NJ 07102-1982, U.S.A. \\
\and
Gavin Lynch\thanks{This article is mainly based on a part of the author's
Ph.D. dissertation at New Jersey Institute of Technology, under the
guidance of Wenge Guo.} \\
Catchpoint Systems, Inc. \\
New York, NY 10003, U.S.A. \\
\and
Joseph P. Romano\thanks{The
research of the author was supported in part by NSF Grant DMS-0707085.} \\
Departments of Statistics and Economics\\
 Stanford University\\
Stanford, CA 94305-4065, U.S.A. \\
}

\date{}

\maketitle

\begin{abstract}
In many large scale multiple testing applications, the hypotheses often have a known graphical structure, such as gene ontology in gene expression data. Exploiting this graphical structure in multiple testing procedures can improve  power as well as aid in interpretation. However, incorporating the structure into large scale testing procedures and proving that an error rate, such as the false discovery rate (FDR), is controlled can be challenging. In this paper,
we introduce a new general approach for large scale multiple testing, which can aid in developing new procedures under various settings with proven control of desired error rates. This approach is particularly useful for developing FDR controlling procedures, which is simplified as the problem of developing per-family error rate (PFER) controlling procedures. Specifically, for testing hypotheses with a directed acyclic graph (DAG) structure, by using the general approach, under the assumption of independence, we first develop a specific PFER controlling procedure and based on this procedure, then develop a new FDR controlling procedure, which can preserve the desired DAG structure among the rejected hypotheses. Through a small simulation study and a real data analysis, we illustrate nice performance of the proposed FDR controlling procedure for DAG-structured hypotheses.

\end{abstract}

\noindent KEY WORDS: Bonferroni procedure, BH procedure, DAG, false discovery rate, multiple testing, per-family error rate.

\section{Introduction}

In many multiple testing applications, the hypotheses have a known graphical structure such as gene ontology  in gene expression data. However, there are few testing procedures controlling large-scale error rates, such as the false discovery rate (FDR), that account for the graphical structure in the hypotheses. In this paper, we  look at testing hypotheses with a directed acyclic graph (DAG) structure (though the theory applies more generally).

Testing on a complex structure such as a DAG has applications to the testing gene ontology terms for phenotype association (Goeman and Mansmann, 2008) and clinical trials (Dmitrienko et al., 2007; Dmitrienko and Tamhane, 2013). Moreover, a DAG structure is a very generic structure and can take the form of a hierarchy (Lynch and Guo, 2016; Meinshausen, 2008) or a fixed sequence (Bauer et al., 1998; Lynch et al., 2017). We are motivated to study the DAG structure since it is a very general structure and to our knowledge, there are only a few procedures exist that exploit the structure in order to  control a large scale error rate such as the FDR (Liang and Nettleton, 2010;
Ramdas et al., 2017).

With regard to procedures that test hypotheses along a DAG, there has been some work for procedures controlling the family-wise error rate (FWER). Goeman and Mansmann (2008) proposed a FWER controlling procedure for testing gene ontology terms called the \emph{focus level} method which preserves the graphical structure of gene ontology. The procedure starts by testing a pre-specified subset of the hypotheses. At each step, the procedure expands the set of tested hypotheses based on the rejections in the previous step. Dmitrienko et al. (2007) and Dmitrienko and Tamhane (2013) proposed general FWER controlling methods for testing hierarchically ordered hypotheses with applications to clinical trials. The hypotheses are organized into families and the families are tested sequentially. A decision regarding whether a hypothesis can be tested is made based on which hypotheses are rejected in the previously tested families. For recent works on developing FWER controlling methods for testing hypotheses structured as a tree or a graph, see Goeman and Finos (2012), Klinglmueller et al. (2014), and Meijer and Goeman (2015, 2016).

Developing procedures that account for the graphical structure in the hypotheses can be challenging. Furthermore, proof of control of complex large-scale error rates such as the FDR is often more difficult than simpler error rates such as the FWER or the per-family error rate (PFER). In this paper, we introduce a new general approach (Theorem 3.1) for developing procedures controlling large-scale error rates, including the FDR. In the case of the FDR, we show that there exists a close connection between the PFER and the FDR, and that in many cases a new FDR controlling procedure can be developed simply by considering a corresponding PFER controlling procedure. Hence, our proposed framework greatly facilitates the development of new multiple testing procedures.

In particular, a new procedure (Theorem 5.1) for testing hypotheses which have a DAG structure is developed, accompanied by theory of  FDR control. Specifically, our procedure only rejects a hypothesis if all of its parent hypotheses are rejected. The advantages of accounting for the underlying structure of the hypotheses are two-fold. First, there is a potential gain in power in the testing of these hypotheses. Indeed, our simulation study and real data analysis show that our proposed procedure compares favorably against the BH procedure in terms of power by accounting for the underlying DAG structure. Second, the rejected hypotheses maintain their hierarchical integrity,  which can enhance interpretation. That is, the rejected hypotheses form a DAG that is a subset of the tested DAG. With regard to testing hypotheses along a DAG, we also show that the BH procedure can be slightly modified so that the rejected hypotheses maintain their hierarchical integrity. This modification reduces the power of the usual BH procedure but benefits from the fact that the rejection set preserves the DAG structure.

The rest of this paper is organized as follows. Section 2 introduces some basic notation and concepts.
Section 3 describes a general approach for developing multiple testing methods controlling various error rates.
Section 4 describes a PFER controlling procedure for testing hypotheses with a DAG structure and in Section 5, we use our proposed approach in Section 3 to turn this PFER controlling procedure into a FDR controlling procedure. A small simulation study  is presented in Section 6 and an empirical data analysis  is given in Section  7. Finally, some concluding remarks are given in Section 8 and all proofs are deferred to Section 9.

\section{Preliminaries}
Suppose that $H_i, i = 1, \ldots, m$, are the $m$ null hypotheses to be tested simultaneously based on their respective $p$-values
 $P_i, i = 1, \ldots, m$. Let $m_0$ denote the (unknown) number of  true null hypotheses.  The marginal distributions of the true null $p$-values are assumed to satisfy the following basic assumption:
\begin{equation}
\pr{P_i \le p} \le p \text{~~~for~any~} p \in (0, 1) \text{~when~} H_i \text{~is~true}. \label{EQN_UNIFORM}
\end{equation}
Regarding the joint dependence of the $p$-values, we mainly focus on independence and positive dependence. The typical positive dependence assumption used in multiple testing literature is the PRDS property, which is often satisfied in some typical multiple testing situations (Benjamini and Yekutieli, 2001; Sarkar, 2002).

\begin{assumption} \label{ASM_POS_DEPENDENCE} \emph{(The PRDS property)}. For any coordinatewise non-decreasing function of the p-values $\psi$,
\begin{equation}
\expv{\psi(P_1, \dots, P_m) \cond P_i \le p} \text{ is non-decreasing in $p$ for each $i$ such that $H_i$ is true}.
\end{equation}
\end{assumption}

For a given multiple testing procedure, let $V$ be the number of rejected true null hypotheses among $R$ rejected null hypotheses. Then, the per-family error rate (PFER), the familywise error rate (FWER), and false discovery rate (FDR) of this procedure are defined respectively as $\text{PFER} = \expv{V}$, $\text{FWER} = \pr{V > 0}$, and $\text{FDR} = \expv{\frac{V}{R \vee 1}}$, where $R \vee 1 = \max\{R, 1\}$. Let us first review some common PFER controlling procedures, which are often used for controlling the FWER.   Indeed, control of the PFER implies control of the FWER (but not conversely). We assume each of the procedures is applied at a pre-specified level $\alpha$.

\begin{exmp}\label{Bonf}\rm (The Bonferroni Procedure). For each $i = 1, \dots, m$, the Bonferroni procedure rejects $H_i$ if $P_i \le \alpha/m$. The Bonferroni procedure controls the PFER under arbitrary dependence. A variant of the Bonferroni procedure is the weighted Bonferroni procedure. Given a sequence of weights $w_1, \dots, w_m$ such that $\sum_{i=1}^{m}w_i = 1$, the weighted Bonferroni procedure rejects $H_i$ if $P_i \le w_i\alpha/m$. Another variant is the oracle Bonferroni procedure which rejects $H_i$ if $P_i \le \alpha/m_0$. The oracle Bonferroni procedure assumes that $m_0$ is known in advance. The weighted and oracle Bonferroni procedures also control the PFER under arbitrary dependence. \qed
\end{exmp}

\begin{exmp}\label{adaptive Bonf} \rm (The Adaptive Bonferroni Procedure). Given $\gamma \in (0,1)$, for each $i = 1, \dots, m$, the adaptive Bonferroni procedure rejects $H_i$ if $P_i \le \alpha/\widehat{m}_0$ where $\widehat{m}_0 = (\sum_{i=1}^{m}\ind{P_i > \gamma} +1)/(1-\gamma)$. The adaptive Bonferroni procedure controls the PFER under independence. \qed
\end{exmp}

Many multiple testing procedures are described as  stepup procedures, which have the following form.  Given a sequence of non-decreasing critical constants $\alpha_1 \le \cdots \le \alpha_m$, a stepup procedure orders the $p$-values: $P_{(1)} \le \cdots \le P_{(m)}$ with corresponding hypotheses $H_{(1)}, \dots, H_{(m)}$. Hypotheses $H_{(1)}, \dots, H_{(R)}$ are rejected and the rest are accepted where $R = \max\{0 \le r \le m: P_{(r)} \le \alpha_r\}$ with $P_{(0)} \equiv 0$. Our proposed approach and  the developed methods  are motivated by such stepup procedures.

\section{A General Approach}

In this section, we introduce a general approach for developing complex multiple testing procedures controlling various error rates based on simple existing procedures. Specifically,
 new FDR controlling procedures are developed  by utilizing simple PFER controlling procedures.

We motivate the general approach by  first studying the relationship between the Bonferroni procedure and the BH procedure in the example below.

\begin{exmp} \rm
Recall that the popular BH procedure is the stepup procedure with critical constants  $i\alpha/m, i = 1, \dots, m$ (Benjamini and Hochberg, 1995). The number of rejections by the BH procedure is $R = \max\{0 \le r \le m: P_{(r)} \le r\alpha/m\}$. For any given $0 < \beta < 1$, let $R_{bonf}(\beta) = \sum_{i=1}^{m}\ind{P_i \le \beta/m}$ be the number of rejections by the Bonferroni procedure applied at level $\beta$. Since the event $\{P_{(r)} \le r\alpha/m\}$ is equivalent to the event $\{r \le R_{bonf}(r\alpha)\}$, we see that the number of rejections $R$ by the BH procedure can equivalently be expressed as
\begin{equation}
R = \max\{0 \le r \le m: r \le R_{bonf}(r \alpha)\}. \label{EQN_BH}
\end{equation}
Thus, the Bonferroni procedure applied at level $R\alpha$ is equivalent to the BH procedure applied at level $\alpha$. \qed
\end{exmp}

Notice that (\ref{EQN_BH}) does not determine the number of rejections based on the $p$-values directly but instead determines the number of rejections from the Bonferroni procedure. We can readily generalize this idea for other procedures, not just the Bonferroni procedure.

For a multiple testing procedure, let $L$ denote a simple measure of type 1 errors when testing all the $m$ hypotheses. For convenience, we always assume $L = 0$ if $V = 0$. Typically, we will set $L = V$ so that $L$ is the total number of type 1 errors. In some cases, we will also consider, with less emphasis, other error measures such as $V\ind{V \ge k}$, where $I$ is indicator function and $k$ a given positive integer. Suppose that the expected value of $L$, $E(L)$, is used as a simple type 1 error rate and there is some procedure available that controls $\expv{L}$ at nominal level $\alpha$;
the procedure satisfies  $\expv{L} \le \alpha$.  We also assume that the procedure satisfies the property of $\alpha$-monotonicity, i.e., the number of rejections by the procedure is a non-decreasing function of the level $\alpha$. We will refer to this procedure as the {\it base} procedure. For example, if $L = V$, then a PFER controlling procedure could be a base procedure.

The  main objective in this section is to control a general two-parameter error rate by basing its construction on $L$.
To describe the two-parameter error rate,  suppose that  $W:\{0, 1, \ldots, m\} \rightarrow [0, 1]$ is a non-increasing weight function of the number of rejections $R$. Then, the two-parameter error rate that we are interested in  controlling takes the general form
\begin{equation}
\expv{W(R) L}. \label{EQN_L_WEIGHTED}
\end{equation}
For example, let $L = V$. If $W(R) = 1$, then (\ref{EQN_L_WEIGHTED}) is simply the PFER. If $W(R) = 1/\maxR$, then (\ref{EQN_L_WEIGHTED}) is the popular FDR. We will give particular attention to the latter case.

We are interested in controlling (\ref{EQN_L_WEIGHTED}) at a fixed, pre-specified level $\alpha$. We do this by determining a value $\beta$, which is based on the $p$-values, such that the base procedure is applied at level $\beta$, i.e., the procedure rejects hypotheses according to rejection thresholds that are determined based on level $\beta$. Given a weight function $W$ and a base procedure, a general large-scale testing procedure (GELS) aiming at controlling (\ref{EQN_L_WEIGHTED}) is defined as follows.
\begin{definition}\label{GELS procedure} \emph{(The GELS Procedure).} Let $R_{base}(\beta)$ be the number of rejections by the base procedure applied at level $\beta$.
\begin{enumerate}
\item Let \begin{equation}
R = \max\left\{r \in \{0, \dots, m\}: r \le R_{base}(\alpha/ W(r))\right\}. \label{EQN_SINGLE_FAMILY_R}
\end{equation}
\item Test the hypotheses $H_1, \dots, H_m$ by applying the base procedure at level $\alpha/W(R)$. 
\end{enumerate}
\end{definition}

\begin{remark}\rm \label{REMARK_SELF_CONSISTENCY}
It should be noted that $R$ in (\ref{EQN_SINGLE_FAMILY_R}) is equal to the number of rejections by the base procedure at level $\alpha/W(R)$. That is, $R = R_{base}(\alpha/W(R)).$ In the literature, this property has been referred to as self-consistency, see Blanchard and Roquain (2008). To see why this property holds, we note by (\ref{EQN_SINGLE_FAMILY_R}) that $R \le R_{base}(\alpha/W(R))$ and $R+1 > R_{base}(\alpha/W(R+1))$, which implies $R \ge R_{base}(\alpha/W(R+1)) \ge R_{base}(\alpha/W(R)) \ge R$, where the second inequality follows from the fact that $R_{base}(\alpha/W(r))$ is a non-decreasing function of $r$. Thus, $R = R_{base}(\alpha/W(R))$. ~\qed
\end{remark}

\begin{remark}\rm \label{REMARK_STEPUP}
It is easy to see that if $W = 1/\maxR$ and the base procedure is the Bonferroni procedure, then (\ref{EQN_SINGLE_FAMILY_R}) reduces to (\ref{EQN_BH}). In fact, the GELS procedure reduces to a stepup procedure when  the base procedure is any single step procedure, not just the Bonferroni procedure. Indeed, suppose the base procedure is a single step procedure whose critical constant is the non-decreasing function $t: \mathbb{R}^+ \rightarrow \mathbb{R}^+$ such that given the significance level $\beta$, $H_i$ is rejected if $P_i \le t(\beta)$. Then, (\ref{EQN_SINGLE_FAMILY_R}) reduces to $R = \max\left\{r \in \{0, \dots, m\}: P_{(r)} \le t(\alpha/ W(r))\right\}$ so that the GELS procedure reduces to a stepup procedure with critical constants $t(\alpha/W(i))$ for $i = 1, \ldots, m$. ~\qed
\end{remark}

\begin{remark}\rm
It is straightforward to find $R$ in (\ref{EQN_SINGLE_FAMILY_R}) by using a simple algorithm below:
\begin{enumerate}
\item Initialize $r_1 = m$ and $t = 1$.
\item Let $r_{t+1}$ be the number of rejections by applying the base procedure at level $\alpha/W(r_t)$.
\item If $r_{t+1} = r_t$, then set $R = r_t$ and stop. Otherwise, increase $t$ by 1 and repeat step 2.
\end{enumerate}
It is clear  that the above algorithm  to determine the number of rejections  is similar in spirit  to  a stepup procedure. ~\qed
\end{remark}

Since the main goal in this section is to control the overall error rate (\ref{EQN_L_WEIGHTED}), let us describe the type of base procedures which are used to control $E(L)$. The error measure $L$ can be classified as two cases: the binary case and the non-binary case.  For example, if $L = \ind{V>0}$, which is binary,  then $\expv{L}$ is the usual FWER.  On the other hand, if $L = V$, which is non-binary, then $\expv{L}$ is the PFER. We will discuss the binary case first where $L$ can only take on $0$ and a positive value.

Given a binary error measure $L(\alpha)$ of the base procedure at level $\alpha$, we introduce a specific assumption of positive dependence based on $L$ as follows.
\begin{assumption} \label{ASM_POS_DEPENDENCE_L} \emph{(Positive Dependence).}
Given a binary error measure $L$ and a base procedure, for any coordinatewise non-decreasing function of the $p$-values $\psi$,
\begin{equation}
\expv{\psi(P_1, \dots, P_m) \cond L(\alpha) > 0} \text{ is non-decreasing in $\alpha$}.
\end{equation}
\end{assumption}

Note that the conventional positive dependence assumption, Assumption \ref{ASM_POS_DEPENDENCE}, is equivalent to the condition that Assumption \ref{ASM_POS_DEPENDENCE_L}
is satisfied for each $L_i(\alpha) = \ind{P_i \le \alpha}, i = 1, \dots, m$ such that $H_i$ is true.
If the $p$-values are independent, then we have the following result.

\begin{lemma} \label{LEMMA_POS_INDPENDENCE}
Suppose the binary error measure $L$ is of the form $L(\alpha) = \ind{ \bigcap_{i=1}^{m}P_i \le t_i(\alpha)}$ for non-decreasing functions $t_i: \mathbb{R}^+ \rightarrow \mathbb{R}^+, i = 1, \dots, m$. If the $p$-values are independent, then Assumption \ref{ASM_POS_DEPENDENCE_L} is satisfied.
\end{lemma}

\begin{remark}\rm
Assumption \ref{ASM_POS_DEPENDENCE_L} is weaker than the (PosDep) assumption given in Delattre and Roquain (2015), which is a multivariate version of the PRDS property (Benjamini and Yekutieli, 2001) and its slight variant (Sarkar and Guo, 2010), and the (PosDep) assumption is implied by the $\text{MTP}_2$ condition (Shaked and Shanthikumar, 1994). Thus, Assumption \ref{ASM_POS_DEPENDENCE_L} is also satisfied under the (PosDep) assumption or
the $\text{MTP}_2$ condition.
\end{remark}

Generally, binary error measures are of limited interest.  However, they provide a good starting point for studying non-binary error measures, which is the focus for the remainder of this section. In particular, we will consider a general  non-binary error rate $L$ and a base procedure controlling $E(L)$  satisfying the following three conditions:

\medskip

\textbf{Condition C1:} $L$ is less than or equal to the sum of $s$ binary error rates, $L_1, \dots, L_s$, each of which is a non-increasing function of the $p$-values, where $s$ is a positive integer.

\medskip

\textbf{Condition C2:} $\expv{L_i(\alpha)} \le c_i\alpha$ for every fixed $\alpha > 0$, where $c_1, \dots, c_s$ are non-negative constants such that $\sum_{i=1}^{s}c_i \le 1$.

\medskip

\textbf{Condition C3:} Assumption \ref{ASM_POS_DEPENDENCE_L} is satisfied for each $L_i, i = 1, \dots, s$.

\begin{remark}\rm
Condition C1 depends on the base procedure and $L_i(\cdot)$ is a function of the pre-specified level $\alpha$ and the $p$-values. Generally, the condition C1 can be regarded as a property of decomposability for the error measure $L$, condition C2 as a property of local error rate control for the base procedure, and condition C3 as a property of the joint dependence of the
$p$-values. Also, if conditions C1 and C2 are satisfied, then it must be the case that $\expv{L(\alpha)} \le \alpha$. Indeed, $\expv{L(\alpha)} \le \sum_{i=1}^{s}\expv{L_i(\alpha)} \le \sum_{i=1}^{s}c_i\alpha \le \alpha$. ~\qed
\end{remark}

Consider the non-binary error measure $L = V$, the total number of falsely rejected hypotheses at level $\alpha$. It is easy to see that the type 1 error measure $L$ of the conventional Bonferroni procedure can be broken into $m$ binary error rates, $L_i(\alpha/m) = \ind{P_i \le \alpha/m, H_i \text{ is true}}, i = 1, \dots, m$. Let $c_i = 1/m$ for $i = 1, \dots, m$, then the Bonferroni procedure satisfies conditions C1 and C2. Moreover, if the $p$-values satisfy Assumption \ref{ASM_POS_DEPENDENCE}, then condition C3 is also satisfied. It is interesting to note that not all PFER controlling procedures satisfy conditions C1 and C2. We will present an example of such a procedure later in this section.

\begin{theorem} \label{THM_COMPOSITE}
Given a type 1 error measure $L$ and a base procedure controlling $E(L)$, if conditions C1-C3 are satisfied, then the GELS procedure  defined in Definition \ref{GELS procedure}  controls the two-parameter error rate  defined in (\ref{EQN_L_WEIGHTED});  that is,    $\expv{W(R)L} \le \alpha$.
\end{theorem}

Let us consider three examples to show why Theorem \ref{THM_COMPOSITE} is useful.
\begin{exmp} \rm \label{EXMP_ADAPT_FDR}
In this example, we will show how Theorem \ref{THM_COMPOSITE} can be used for developing FDR controlling procedures. Let $L = V$ and $W = 1/\maxR$. For the base procedure controlling $E(L)$, consider the adaptive Bonferroni procedure described in Example \ref{adaptive Bonf}. In the following, we will show that conditions C1-C3 are satisfied under independence. Define
$$L_i = \ind{P_i \le \alpha/\widehat{m}^{(-i)}_0, H_i \text{ is true}},$$
where
$$\widehat{m}^{(-i)}_0 = (\sum_{j \neq i}\ind{P_j > \gamma} +1)/(1-\gamma).$$
Thus,
\[
\sum_{i = 1}^{m}L_i \ge \sum_{i=1}^{m}\ind{P_i \le \alpha/\widehat{m}_0, H_i \text{ is true}} = L.
\]
So condition C1 is satisfied. Secondly, when $H_i$ is true, it is shown in Guo (2009) that under independence, $\expv{L_i(\alpha)} \le \alpha/m_0$. Setting $c_i = 1/m_0$ if $H_i$ is true and 0 otherwise, we see that condition C2 is satisfied. Finally, by Lemma \ref{LEMMA_POS_INDPENDENCE} and the double expectation theorem, it is easy to see that condition C3 is satisfied under independence. Therefore, by Theorem \ref{THM_COMPOSITE}, the GELS procedure based on the adaptive Bonferroni procedure strongly controls the FDR at level $\alpha$ under independence.

It should be noted that the above GELS procedure is equivalent to the adaptive BH procedure introduced in Storey et al. (2004) (also see Benjamini et al., 2006; Sarkar, 2008), which is defined below. Define
$$R = \max\{1 \le r \le m: P_{(r)} \le r\alpha/\widehat m_0\}.$$
Then, reject $H_i$ iff $P_i \le R\alpha/\widehat m_0$ for any $i = 1, \ldots, m$. ~\qed
\end{exmp}

\begin{exmp}\rm \label{EXMP_WEIGHTED_FDR}
Other base procedures, such as the weighted Bonferroni procedure and the oracle Bonferroni procedure described in Example \ref{Bonf}, could also be used with the general GELS procedure for developing FDR controlling procedures. If the weighted Bonferroni procedure is used as the base procedure, then the resulting
GELS procedure is equivalent to the weighted BH procedure (Genovese et al., 2006), which is defined below. Let $P^*_{(1)} \le \cdots \le P^*_{(m)}$ denote the ordered version of the weighted $p$-values, $P_1/W_1, \ldots, P_m/W_m$, and define
$$R^* = \max\{1 \le r \le m: P^*_{(r)} \le r\alpha/m\}.$$
Then, reject $H_i$ iff $P_i/W_i \le R^*\alpha/m$ for any $i = 1, \ldots, m$.

If the oracle Bonferroni procedure is used as the base procedure, then the resulting
GELS procedure is equivalent to the oracle BH procedure, which is defined below. Define
$$\tilde{R} = \max\{1 \le r \le m: P_{(r)} \le r\alpha/m_0\}.$$
Then, reject $H_i$ iff $P_i \le \tilde{R}\alpha/m_0$ for any $i = 1, \ldots, m$. ~\qed
\end{exmp}

\begin{exmp}\rm \label{EXMP_GENERALIZED_FDR}
Theorem \ref{THM_COMPOSITE} can also be used to develop new multiple testing procedures for controlling other two-parameter error measures other than the FDR. For example, consider the generalized false discovery rate ($k$-FDR), which is $\expv{V\ind{V \ge k}/(R \vee k)}$ for some pre-specified positive integer $k$ (Sarkar, 2007; Sarkar and Guo, 2009). Here, we have $L = V\ind{V \ge k}$ and $W = 1/(R \vee k)$. First, we will describe a base procedure controlling $E(L)$, and then we will show that conditions C1-C3 are satisfied under independence. The chosen base procedure is the single step procedure with critical constant $t = [(k-1)! \alpha / \binom{m}{k}]^{1/k}$. To show conditions C1-C3 are satisfied, let $I_0$ denote the index set of true null hypotheses and $\mathcal{C}_i^0 = \{ \mathcal{J} : \mathcal{J} \subseteq I_0 \text{ such that } |\mathcal{J}| = i \}$ for any $i=1, \ldots, m_0$. Let $L_\mathcal{J} = \ind{\max_{j \in \mathcal{J}}P_j \le t}/(k-1)!$ and define $c_\mathcal{J} = 1 /\binom{m}{k}$ for any  $\mathcal{J} \in \mathcal{C}_k^0$ and $k=1, \ldots, m_0$. Then, the following two inequalities hold, which are proved in Section 9.3:
\begin{equation}
\sum_{\mathcal{J} \in \mathcal{C}_k^0}L_\mathcal{J} \ge L ~\text{ and }~
\sum_{\mathcal{J} \in \mathcal{C}_k^0}c_\mathcal{J} = \binom{m_0}{k}/\binom{m}{k} \le 1 \label{EQN_KFDR}
\end{equation}
for any $k=1, \ldots, m_0$. Hence, conditions C1  and C2 are satisfied.  Finally, by Lemma \ref{LEMMA_POS_INDPENDENCE}, condition C3 is also satisfied under independence.

Thus, by Theorem \ref{THM_COMPOSITE}, the GELS procedure based on the above single step method controls the $k$-FDR at level $\alpha$ under independence. Furthermore, by Remark \ref{REMARK_STEPUP}, this GELS procedure is equivalent to the stepup procedure with critical constants $\alpha_i$ satisfying
\begin{equation}
\alpha_i = \left[(k-1)! \max(k, i)\alpha \big/ \binom{m}{k}\right]^{1/k}, i = 1, \dots, m.
\end{equation}
It should be noted that this stepup procedure has different critical constants than the stepup $k$-FDR controlling procedure in Sarkar (2007) for independent $p$-values. Although neither critical constants dominates the other for these two stepup procedures, the first few critical constants (corresponding to $i = 1, 2, 3$, etc.) of our procedure tend to be much larger than those of Sarkar's procedure, while the last few critical constants of Sarkar's procedure tend to be much larger than those of our procedure. ~\qed
\end{exmp}

\begin{remark}\rm \label{REM_LOSE_FDR}
In general, if conditions C1-C3 are not satisfied, then the GELS procedure might lose control of the error rate $\expv{W(R) L}$. Specifically, for  FDR control, a positive dependence condition (which corresponds to condition C3) is typically required for the $p$-values. If the condition is not satisfied, even for the BH procedure, it can lose control of the FDR (Guo and Rao, 2008). In this remark, we will provide an example below that shows that condition C2 is also necessary, i.e., it is not enough for the base procedure to control $\expv{L}$ without satisfying condition C2.

Suppose $m=2$ and both hypotheses $H_1$ and $H_2$ are true.  Assume the corresponding independent $p$-values $P_1$ and $P_2$ are  each uniform on $(0,1)$.  Let $L = V$, and consider the following base procedure: Reject $H_1$ if $P_1 \le \alpha/(1+\alpha)$ and reject $H_2$ if $H_1$ is rejected and $P_2 \le \alpha$. It is easy to see that conditions C1 and C3 are satisfied and the PFER is controlled under independence:
\begin{align*}
    &\expv{V}
=    \pr{P_1 \le \frac{\alpha}{1+\alpha}} + \pr{P_1 \le \frac{\alpha}{1+\alpha}, P_2 \le \alpha}
\le  \frac{\alpha}{1+\alpha} + \frac{\alpha^2}{1+\alpha}
=    \alpha.
\end{align*}
Setting $W = 1/\maxR$, the GELS procedure applies this base procedure at level $R\alpha$.
In this case, the FDR is the probability of any rejection, so
 $$ \expv{W L} = FDR = \pr{ R > 0 } = \pr{R=2} + \pr{R=1}~.$$
 But
 $$\pr{R=2} = \pr{ P_1 \le \frac{2 \alpha}{1 + 2 \alpha} , P_2 \le 2 \alpha } = \frac{2 \alpha}{1 + 2 \alpha} \cdot 2 \alpha$$
 and
 $$\pr{R=1} = \pr{ P_1 \le \frac{\alpha}{1 + \alpha} , P_2 > 2 \alpha } = \frac{ \alpha}{1+ \alpha} ( 1- 2 \alpha )~.$$
 Summing these two expressions yields
 $$ \expv{W L} = \frac{ \alpha ( 4 \alpha +  1)}{(1 + 2 \alpha )(1 + \alpha )} > \alpha$$
 if $\alpha <  1/2$.
To see why this procedure does not satisfy condition C2, note that $V$ can be broken up into binary error rates $L_1 = \ind{H_1 \text{ rejected}}$ and $L_2 = \ind{H_2 \text{ rejected}}$. The above base procedure controls $\expv{L_1}$ at $\alpha/(1+\alpha)$ and $\expv{L_2}$ at $\alpha^2/(1+\alpha)$. However, $c_1 = c_2 = 1$ are the only constants such that $\expv{L_1(\alpha)} \le c_1\alpha$ and $\expv{L_2(\alpha)} \le c_2\alpha$ for every $\alpha > 0$. Thus, the procedure does not satisfy condition C2. ~\qed
\end{remark}

\begin{remark}\rm
Mekaldji et al. (2016) consider a variant of the FDR, penalized false discovery rate $E(W(R)V)$, which can be regarded as a special form of $E(W(R)L)$ with $L = V$. Thus, by using our Theorem 3.1, we can easily derive some procedures controlling this error rate under positive dependence. This is interesting, since our main goal is still to develop FDR controlling procedure for testing graphically structured hypotheses. It shows that our general approach is also useful in developing procedures controlling other forms of type 1 error rates.
\end{remark}


\section{Testing Hypotheses with a DAG Structure}

We now consider testing hypotheses organized in a directed acyclic graph (DAG) where the hypotheses are the vertices and a hypothesis is not tested unless all of the hypotheses associated with incoming edges have been rejected. Such a testing strategy has application to  testing of terms in a gene ontology, where it is natural to test a term only if its more general parent terms show significance. Furthermore, this structure is very general and can take the form of a hierarchy (Lynch and Guo, 2016; Meinshausen, 2008), or a fixed sequence (Bauer et al., 1998; Lynch et al., 2017). We will develop a new procedure to test hypotheses in this manner and prove that it controls the PFER. In next section, we will use the procedure to develop a new FDR controlling procedure.

Let $\mathcal{T}_i$ be the set of hypotheses associated with the incoming edges of $H_i$ so that $H_i$ is only tested if each $H_j \in \mathcal{T}_i$ is rejected. That is, $\mathcal{T}_i$ is the set of parent hypotheses of $H_i$. Now, define $\mathcal{T}^{(0)}_i = \{H_i\}$ and $\mathcal{T}^{(k)}_i = \bigcup_{H_j \in \mathcal{T}^{(k-1)}_i}\mathcal{T}_j$ for $k = 1, \ldots, m$.
Then, the set of ancestor hypotheses of $H_i$ and $H_i$ itself is $\mathcal{D}_i = \bigcup_{k=0}^{m}\mathcal{T}_i^{(k)}$. It follows that $H_i$ is rejected if and only if  every hypothesis in $\mathcal{D}_i$ is rejected. Opposite to $\mathcal{D}_i$, let $\mathcal{M}_i$ be the set of descendant hypotheses of $H_i$ and $H_i$ itself so that $\mathcal{M}_i = \{H_j: H_i \in \mathcal{D}_j\}$. We will refer to a leaf hypothesis as any hypothesis that is not a parent hypothesis (i.e. it does not having any outgoing edges). Mathematically, $H_i$ is a leaf hypothesis if $\mathcal{M}_i = \{H_i\}$ or, alternatively, $|\mathcal{M}_i| = 1$. We will typically use $\ind{|\mathcal{M}_i| > 1}$ to indicate that that $H_i$ is not a leaf hypothesis. Let $\ell$ be the total number of leaf hypotheses. In addition, we group the $m$ tested hypotheses into families $\mathcal{F}_1, \dots, \mathcal{F}_m$, where hypothesis $H_i$ is in set $\mathcal{F}_j$ if the longest path to $H_i$ traverses $j-1$ edges. Formally,
\begin{align*}
&\mathcal{F}_1 = \{H_i: \mathcal{T}_i = \emptyset\},\\
&\mathcal{F}_2 = \{H_i: H_i \notin \mathcal{F}_1, \mathcal{T}_i \subseteq \mathcal{F}_1\},\\
&\dots \\
&\mathcal{F}_m = \left\{H_i: H_i \notin \bigcup_{j=1}^{m-1}\mathcal{F}_j, \mathcal{T}_i \subseteq \bigcup_{j=1}^{m-1}\mathcal{F}_j \right\}.
\end{align*}

\begin{figure}
 \centering
  \includegraphics[scale=.5]{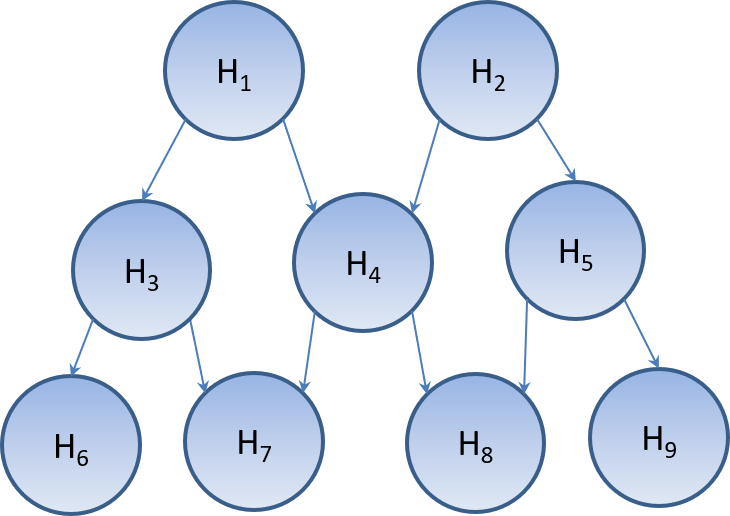}
  \caption{An example of a directed acyclic graph structure with 9 hypotheses. $H_3$ is only tested if $H_1$ is rejected, $H_4$ is only tested if $H_1$ and $H_2$ are rejected, $H_5$ is only tested if $H_2$ is rejected,
$H_6$ is only tested if $H_3$ is rejected, $H_7$ is only tested if $H_3$ and $H_4$ are rejected, $H_8$ is only tested if $H_4$ and $H_5$ are rejected,
and $H_9$ is only tested if $H_5$ is rejected.}
  \label{IMG_DAG_EXAMPLE}
\end{figure}

Consider the example depicted in Figure \ref{IMG_DAG_EXAMPLE}. Here, $\mathcal{T}_4 = \{H_1, H_2\}$, $\mathcal{D}_6 = \{H_1, H_3, H_6\}$,
$\mathcal{M}_4 = \{H_4, H_7, H_8\}$, $\mathcal{F}_1 = \{H_1, H_2\}$, $\mathcal{F}_2 = \{H_3, H_4, H_5\}$, $\mathcal{F}_3 = \{H_6, H_7, H_8, H_9\}$,
$\mathcal{F}_i = \emptyset$ for $i = 4, \ldots, 9$, $H_6$, $H_7$, $H_8$, and $H_9$ are leaf hypotheses, and $\ell = 4$.

The following generic procedure is used to test hypotheses along a DAG.
\begin{definition} (\emph{DAG Testing Procedure}). Let $\alpha_1 , \ldots , \alpha_m$ be critical constants.
\begin{enumerate}
\item For each hypothesis $H_i$  with no parents,  reject $H_i$ if $P_i \le \alpha_i$; otherwise,  do not reject $H_i$.
\item For each $H_i, i = 1, \dots, m, $ that has not been tested and whose parents have all been tested, if all of $H_i$'s parents have been rejected and $P_i \le \alpha_i$, then reject $H_i$. Otherwise, accept $H_i$. Repeat this step until no hypotheses are left to test.
\end{enumerate}
\end{definition}

Before presenting the main result of this section, we introduce one more notation. For each pair of hypotheses, $H_i$ and $H_j$, define $s_{i, j}$ as follows.
\begin{align}
s_{i, j} =
\begin{cases}
0  &\text{if } H_i \notin \mathcal{D}_j, \\
1  &\text{if } H_i = H_j, \\
\sum_{H_k \in \mathcal{T}_j}\dfrac{s_{i,k}}{|\mathcal{T}_j|}  &\text{if } H_i \in \mathcal{D}_j \setminus \{H_j\}.
\end{cases} \label{EQN_S_DAG}
\end{align}
Now, define
\begin{equation} \label{equation:elli}
\ell_i = \sum_{H_j \text{ is a leaf}}s_{i, j}.
\end{equation}
It is easy to see that $\ell_i = 1$ for any leaf hypothesis $H_i$, however, for any non-leaf hypothesis $H_i$, the value of $\ell_i$ depend on the values of $s_{i, j}$. For example, in Figure \ref{IMG_DAG_EXAMPLE}, there are five non-leaf hypotheses $H_i, i = 1, \ldots, 5$ and four leaf hypotheses $H_i, i = 6, \ldots, 9$.
For $i = 1, \ldots, 5$,  the values of $\ell_i$ are respectively $2, 2, \frac{3}{2}, 1$ and $\frac{3}{2}$. Among these values of $\ell_i$, $\ell_1 = 2$ is obtained through
$s_{1, 6} = 1$, $s_{1, 7} = \frac{3}{4}$, $s_{1, 8} = \frac{1}{4}$, and $s_{1, 9} = 0$. For other values of $\ell_i$, they were similarly obtained.

We can interpret $s_{i,j}$ by considering the following analogy. Suppose water flows from each node in the opposite direction of the edges. At each node $H_j$, the amount of water flowing through $H_j$ is divided evenly among its parents so that a proportion of $1/|\mathcal{T}_j|$ of the water flowing through $H_j$ flows through each of $H_j$'s parents. In this analogy, $s_{i,j}$ represents the proportion of water starting from $H_j$ and flowing through $H_i$, and $\ell_i$ represents the total amount of water starting from leaf hypotheses and flowing through $H_i$. If $H_i \notin \mathcal{D}_j$, then no water starting from $H_j$ can reach $H_i$ and thus, $s_{i, j} = 0$. Also, all the water starting from $H_i$ flows through $H_i$ so that $s_{i,i} = 1$. By means of this analogy, we have the following lemma.
\begin{lemma}\label{LEM2}
Let $s_{i, j}$ be defined as in (\ref{EQN_S_DAG}). The following two properties hold.
\begin{itemize}
  \item[(i)] For any fixed $j = 1, \ldots, m$,
  \begin{equation} \label{property1}
  \sum_{H_k \in \mathcal{F}_1}s_{k, j} = 1;
  \end{equation}
  \item[(ii)] For any fixed $i, j= 1, \ldots, m$ such that $i \neq j$,
  \begin{equation} \label{property2}
  s_{i, j} = \sum_{k: H_i \in \mathcal{T}_k}s_{k, j}/|\mathcal{T}_k|.
  \end{equation}
\end{itemize}

\end{lemma}

By using the above analogy, property (i) can be interpreted as: all water starting from $H_j$ must eventually reach the top of the DAG. Property (ii) can be interpreted as: the proportion of water from $H_j$, flowing through one of $H_i$'s children, say $H_k$, and then through $H_i$ is $s_{k, j}/|\mathcal{T}_k|$.

\begin{remark} \rm
In the special configuration where each hypothesis has 0 or 1 parent hypotheses, the directed acyclic graph reduces to the aforementioned hierarchical structure (Lynch and Guo, 2016). In this case, $s_{i, j}$ is 0 or 1 depending on whether $H_i \in \mathcal{D}_j$ and thus, $\ell_i$ is simply the number of leaf hypotheses in $\mathcal{M}_i$. Under such configuration, it is easy to check that the two properties on $s_{i, j}$ in the above lemma holds for any $i, j = 1, \ldots, m$.
\end{remark}

By using the notations $\ell_i$ and $\ell$, we develop a PFER controlling procedures for testing hypotheses with a DAG structure below.

\begin{theorem} \label{THM_PFER_TESTING_PROC}
Let $\ell_i$ be defined as in (\ref{equation:elli}) and $\ell$ the total number of leaf hypotheses.
Given a pre-specified significance level $\alpha$ and a tuning parameter $\lambda > 0$, the DAG testing procedure with critical constants
\begin{align*}
\alpha_i =
\begin{cases}
\min\left(\lambda, \dfrac{\alpha}{\ell}\right)\dfrac{\ell_i}{1+\lambda\ell_i} &\text{ if $H_i$ is not a leaf} \\
\min\left(\lambda, \dfrac{\alpha}{\ell}\right) &\text{ if $H_i$ is a leaf}
\end{cases}
\end{align*}
$i = 1, \dots, m$, controls the PFER at level $\alpha$ under independence.
\end{theorem}

\begin{remark} \rm
It should be noted that the critical constants in Theorem \ref{THM_PFER_TESTING_PROC} rely on a pre-specified tuning parameter $\lambda$. This tuning parameter is necessary due to condition C2 which ensures that $\expv{L_i(\alpha)}$ does not grow too fast with respect to $\alpha$, making it impossible to control the error rate (see Remark \ref{REM_LOSE_FDR}). \qed
\end{remark}

By letting $\lambda = \alpha/\ell$ in Theorem \ref{THM_PFER_TESTING_PROC}, we have the following corollary holds.
\begin{corollary}\label{THM_PFER_TESTING_PROC2}
The DAG testing procedure with critical constants
\begin{align*}
\alpha_i =
\begin{cases}
\dfrac{\alpha}{\ell}\dfrac{\ell_i}{1+\ell_i\alpha/\ell} &\text{ if $H_i$ is not a leaf hypothesis} \\
\dfrac{\alpha}{\ell} &\text{ if $H_i$ is a leaf hypothesis}
\end{cases}
\end{align*}
$i = 1, \dots, m$, controls the PFER at level $\alpha$ under independence.
\end{corollary}

\begin{remark} \rm
In another special configuration, the fixed sequence configuration (Lynch et al., 2017), each hypothesis has 1 parent hypothesis (except $H_1$ which has 0), 1 child hypothesis (except $H_m$ which has 0), and $H_i$ cannot be tested unless $H_1, \dots, H_{i-1}$ have been rejected. Under this configuration, $\ell = 1$ and $\ell_i = 1$ so that the critical constants of the PFER controlling procedure described in Corollary \ref{THM_PFER_TESTING_PROC2} reduce to
\[
\alpha_i = \alpha/(1+\alpha)^{\ind{|\mathcal{M}_i| > 1}}.
\]
Lastly, consider the configuration where all hypotheses are leafs, that is, the hypotheses have no structure, then $l = m$. Under this configuration, the critical constants of this PFER controlling procedure reduce to $\alpha/m$, the critical constant of the conventional Bonferroni procedure. \qed
\end{remark}

It should be noted that the conventional Bonferroni procedure could also be used to test hypotheses along a directed acyclic graph. It is easy to see that such a procedure would control the PFER:
\[
\expv{V} = \sum_{i = 1}^{m}\pr{P_i \le \alpha/m, H_i \text{ is true}, H_i \text{ can be tested}} \le \sum_{i=1}^m \alpha/m = \alpha.
\]
However, for most DAG configurations, the Bonferroni critical constants $\alpha_i = \alpha/m$ are much smaller than those in the PFER controlling procedures described in Theorem \ref{THM_PFER_TESTING_PROC} and Corollary \ref{THM_PFER_TESTING_PROC2}. For example, in the fixed sequence configuration, the Bonferroni critical constants are roughly $m$ times smaller than those used in the procedure of Corollary \ref{THM_PFER_TESTING_PROC2}. Only when the hypotheses have no structure, the two sets of critical constants are the same.

\section{Testing on a Directed Acyclic Graph - FDR Control}

In this section, we discuss FDR control for testing graphically structured hypotheses. Specifically, we use the general approach introduced in Theorem \ref{THM_COMPOSITE}, along with the PFER controlling procedure described in Theorem \ref{THM_PFER_TESTING_PROC} to develop a new FDR controlling procedure to test hypotheses along a DAG. The FDR control of this procedure is established by showing that conditions C1-C3 are satisfied.

\begin{theorem} \label{THM_DAG_TESTING_PROC}
The GELS procedure with $L = V$ and the DAG testing procedure described in Theorem \ref{THM_PFER_TESTING_PROC} used as the base procedure, controls $\expv{W(R)L}$ at level $\alpha$ under independence. Specifically, if we set weight $W(R) = 1/\maxR$, then the GELS procedure controls the FDR at level $\alpha$ under the same condition.
\end{theorem}

We will refer to the above FDR controlling procedure in Theorem \ref{THM_DAG_TESTING_PROC} as the DAG GELS procedure. Its rejection threshold for testing $H_i$ is
\[
\min\left(\lambda, \dfrac{R\alpha}{\ell}\right)\dfrac{\ell_i}{1+\lambda \ell_i\ind{|\mathcal{M}_i| > 1}}
\]
where $R$ is determined from (\ref{EQN_SINGLE_FAMILY_R}).

\begin{remark} \rm
From the above rejection threshold, it is easy to see that for any non-leaf hypothesis $H_i$, it is not rejected by the procedure if $P_i$ is larger than $\lambda\ell_i /(1+\lambda\ell_i)$. This is true no matter how large $R$ is.
Therefore, the value of $\lambda$ affects the performance of this procedure and it is important to choose an appropriate value for $\lambda$ when using this procedure in practice. From an interpretation standpoint, it makes little sense to choose a value of $\lambda$ much larger than $\alpha$ as this implies it is acceptable to reject a hypothesis whose $p$-value is much large than $\alpha$. On the other hand, we do not want to choose a value much smaller than $\alpha$ as this would impose too much power loss. Hence, a value close to $\alpha$ makes the most sense and we have found $\lambda = 2\alpha$ tends to give good results in terms of the number of rejections. \qed
\end{remark}

\begin{remark} \rm
We point out that in both situations where the tests for parent hypotheses are 1) the combinations of the tests of their child hypotheses or 2) univariate tests based on different data, the GELS procedure defined in Theorem \ref{THM_DAG_TESTING_PROC} works.
\end{remark}

As discussed in Section 4, the DAG structure can take on several different forms. Since the DAG structure generalizes a hierarchical structure, our procedure can be used to control the FDR for testing hierarchically ordered hypotheses. In the hierarchical case, our procedure is somewhat similar to the hierarchical testing procedures introduced in Lynch and Guo (2016). Their critical constants are also a function of the number of leaf hypotheses under a parent hypothesis.

In the fixed sequence case, where $\ell = 1$ and $\ell_i = 1, i = 1, \dots, m$, our rejection thresholds for $H_i$ have the following form:
\[
\dfrac{\min(\lambda, R\alpha)}{1+{\lambda\ind{|\mathcal{M}_i| > 1}}}.
\]
Lynch and Guo (2016) introduced a similar FDR controlling procedure for testing hypotheses with a   fixed sequence structure. One major difference between our procedure and the procedures developed in Lynch et al. (2017) and Lynch and Guo (2016) is that their procedures can be implemented in the settings of online data collection and testing, where all the $p$-values are not available while sequential testing (since the data is sequentially collected). However, our procedure, just like conventional $p$-value based stepwise procedures such as the BH procedure, assumes that all the $p$-values are known before testing begins.

Now consider the case where there is no structure in the hypotheses, which is also a special case of the DAG structure. In this case, $\ell = m$, $\ell_i = 1, \dots, m$, and all hypotheses are leaf hypotheses. The rejection threshold of our procedure reduces to $R\alpha/m$ if $\lambda \ge \alpha$, which is the same as that of the conventional BH procedure.

For the sake of comparison, we also consider the BH procedure. Just like the Bonferroni procedure, the set of rejected hypotheses by the BH procedure does not preserve the underlying structure. Also like the Bonferroni procedure, the BH procedure can easily be modified so the hypotheses are tested along a DAG and the structure is preserved. Let us consider a GELS procedure with $L = V$ and $W(R) = 1/\maxR$ described in Section 3. For its base procedure, we consider the DAG testing procedure using simple Bonferroni critical constants as described in the last part of Section 4. By Theorem \ref{THM_COMPOSITE}, the GELS procedure using this DAG testing procedure controls the FDR at level $\alpha$ under certain condition and reduces to a modified BH procedure, which preserves the DAG structure. It should be noted that the modified BH procedure will be less powerful than the conventional BH procedure since the conventional BH procedure is not restricted to testing along a DAG.

\begin{proposition} \label{PROP_DAG_BH}
The GELS procedure with $L = V$ and $W(R) = 1/\maxR$ using the DAG testing procedure with Bonferroni critical constants $\alpha_i = \alpha/m, ~i = 1, \dots, m$, controls the FDR at level $\alpha$ under Assumption \ref{ASM_POS_DEPENDENCE}.
\end{proposition}

\begin{remark} \rm
We need to emphasize that in this section, we only discussed the global FDR control for testing hypotheses in a DAG. For outer nodes FDR and fixed-depth FDR Yekutieli (2008) introduced, it would also be interesting to develop methods controlling these error rates. Generally, the control of some error rate by a particular method does not generally extend to control of a different error rate. For example, Meijer and Goeman (2016) shows
that the BH procedure applied to DAG-structured hypotheses controls the global FDR but not the outer nodes FDR. In addition, for FWER control, Meijer and Goeman (2016) also discussed an appealing property: FWER control on the entire set of rejected hypotheses implies error control on every
subset of the rejected hypotheses. However, this property is not generally shared by the FDR control.
\end{remark}

\section{A Small Simulation Study}

We numerically evaluate the performance of our proposed procedure in Theorem \ref{THM_DAG_TESTING_PROC}, which we refer to as the DAG GELS procedure, and the procedure in Proposition \ref{PROP_DAG_BH}, which we refer to as the DAG BH procedure, in terms of FDR control and average power by comparing them with the conventional BH procedure. It should be noted that the conventional BH procedure does not respect the DAG structure and can test a hypothesis even if all of its parents have not been rejected.

In order to simulate the performance,  we construct a graph of 3 layers deep with 3,003 hypotheses as follows. The first layer, whose hypotheses have no parent, contains 1,000 hypotheses where each hypothesis has 2 children. The second layer contains 1,001 hypotheses and each hypothesis has two parents and two children. The last layer, whose hypotheses have no children, contains 1,002 hypotheses and each hypothesis has 2 parents. The structure of this graph resembles that of Figure \ref{IMG_DAG_EXAMPLE}. A randomly selected proportion of the leaf hypotheses are set to be true with the remaining set to false. Each non-leaf hypothesis is set to be false only if at least one of its child hypotheses are false; otherwise, it is set to be true.

Next, $m$ normal random variables $X_i, i=1, \ldots, m$ are generated  with covariance matrix $\Sigma$ and mean $\mu_i$ to test the $m$ hypotheses $H_i: \mu_i \le 0$ versus $H_i': \mu_i > 0, i = 1, \dots, m$. The $p$-value for testing each hypothesis is calculated using a one-sample one-sided  $Z$-test.  When $H_i$  is true, we set $\mu_i = 0$. When $H_i$  is false, we set $\mu_i$ to a value that depends on its location in the graph. Specifically, $\mu_i = 3$ if it is in the top layer ($H_i$ has no parents), $\mu_i = 2$ if $H_i$ is in the middle layer ($H_i$ has both parents and children), and $\mu_i = 1$ if $H_i$ is in the last layer ($H_i$ has parents but no children). As for the joint dependence of $X_i$, we consider a common correlation structure where $\Sigma$ has off-diagonal components equal to $\rho$ and diagonal components equal to 1.

We set $\alpha = 0.05$. For each procedure we note the false discovery proportion, which is the proportion of falsely rejected hypotheses among all rejected hypotheses, and the  proportion of rejected false null hypotheses among all false null hypotheses. The data was generated and tested 5,000 times and the simulated values of the FDR and average power were obtained by averaging out the 5,000 values of these two proportions, respectively.

\begin{figure}
\centering
\includegraphics[scale=.75, angle = 270, trim=0 0 20 0]{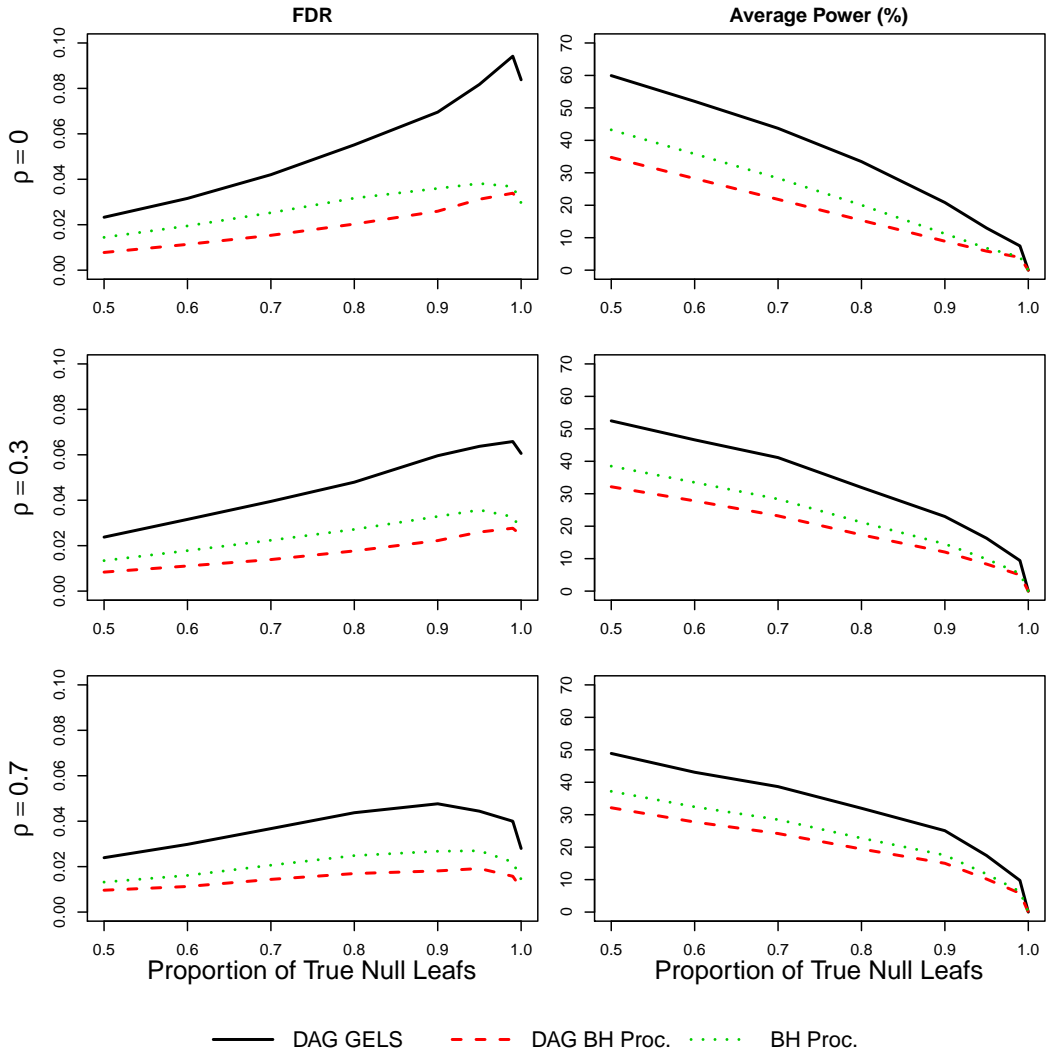}
\caption{Simulated FDR (left column) and average power (right column) of the DAG GELS procedure (solid line), the DAG BH procedure (dashed), and the conventional BH procedure (dotted) under independence ($\rho = 0$, top row), common correlation with $\rho = 0.3$ (middle row), and  common correlation with $\rho = 0.7$ (bottom row).}
\label{IMG_PLOT}
\end{figure}

\begin{figure}
\centering
\includegraphics[scale=.75, angle = 270, trim=0 0 20 0]{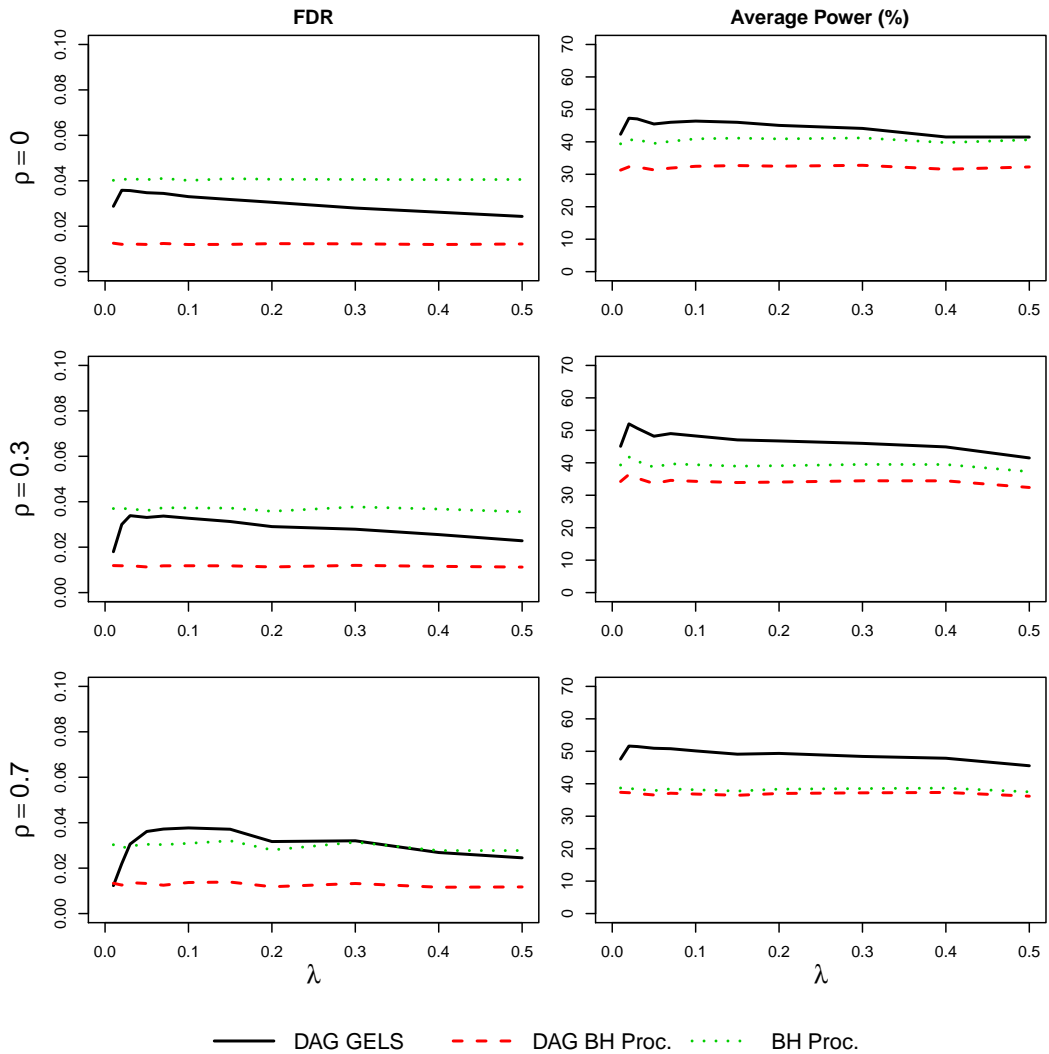}
\caption{Simulated FDR (left column) and average power (right column) of the DAG GELS procedure (solid line), the DAG BH procedure (dashed), and the conventional BH procedure (dotted) under independence ($\rho = 0$, top row), common correlation with $\rho = 0.3$ (middle row), and  common correlation with $\rho = 0.7$ (bottom row). The value of $\lambda$ is varied from 0.01 to 0.5 and the true null leaf proportion is set at 0.9.}
\label{IMG_PLOT2}
\end{figure}

Figures \ref{IMG_PLOT} and \ref{IMG_PLOT2} show the simulated FDR and average power of the DAG GELS procedure, the DAG BH procedure, and the BH procedure under independence ($\rho=0$), common correlation with $\rho = 0.3$, and common correlation with $\rho = 0.7$. For the simulation in Figure \ref{IMG_PLOT}, we fixed $\lambda = 0.1$ and varied the percentage of true null leaf hypotheses. It can be seen in Figure \ref{IMG_PLOT} that all three procedures control the FDR at level 0.05 under the three dependence settings considered. It should be noted that we were only able to show theoretically that the DAG GELS procedure controls the FDR under independence, yet Figure \ref{IMG_PLOT} shows that the DAG GELS procedure controls the FDR under both mild and strong correlation. In terms of power, the DAG GELS procedure outperforms the BH procedure and the DAG BH procedure fairly significantly. It is interesting to note that under independence ($\rho = 0$) and weak dependence ($\rho = 0.3$), the DAG GELS procedure has a smaller FDR than the BH procedure but larger power. This is a desirable but unusual observation for a FDR controlling procedure since it is typically the case that larger power comes at the expense of larger FDR.

For the simulation in Figure \ref{IMG_PLOT2}, we fixed the true null leaf proportion at 0.9 and varied $\lambda$ from 0.01 to 0.5. It can be seen in Figure \ref{IMG_PLOT2} that the FDR is controlled for all values of $\lambda$. The power of the DAG GELS procedure tends to decrease as $\lambda$ increases with the possible exception of very small values of $\lambda$, such as 0.01. Hence, choosing a value of $\lambda$ close to $\alpha$ tends to be powerful and also follows our heuristic guidance mentioned in
Remark 5.1.

\section{Empirical Data Analysis}

We now apply our proposed DAG GELS procedure to the real microarray data set of Golub et al. (1999) involving 27 patients with acute lymphoblastic leukemia and 11 patients with acute myeloid leukemia. The data set is available from the Bioconductor golubEsets package. The data consist of 7,192 expression levels across all 38 patients assayed using Affymetrix Hgu6800 chips. The probes were mapped to gene ontology (GO) biological process terms and of the 7,192 probes, 5,819 were mapped successfully resulting in a total of 10,362 GO terms.

The terms in the GO structure form a directed acyclic graph. If a probe is mapped to a GO term, then it is also mapped to each ancestor of this GO term. Hence, terms lower on the graph refer to more specific gene functions and genes higher on the graph refer to more general gene functions.

Our aim is to determine which GO terms have corresponding expression levels that are different between ALL and AML patients. We formulate a null and alternative hypothesis for each GO term that has at least one probe mapped to it. The null hypothesis states that there is no difference in the expression levels between ALL and AML patients for any of the probes mapped to this term and the alternative states that there is a difference in the expression levels for at least one probe mapped to this term. Hence, if the hypothesis corresponding to a GO term is false, then all of the hypotheses corresponding to its ancestors are also false because the probe with an expression level difference is also mapped to the term's ancestors.

The $p$-value corresponding to each GO term  is calculated using the approach by Goeman et al. (2004).   Hypotheses are tested using the DAG GELS procedure with $\lambda=\alpha, 2\alpha, 4\alpha$ and $10\alpha$, the DAG BH, and the BH procedure, respectively. Table \ref{TABLE_DAG_REJECTIONS} lists the number of rejections for these procedures at various significance levels $\alpha$. We can observe from Table \ref{TABLE_DAG_REJECTIONS} that for all the significance levels, the DAG GELS procedure with different values of
$\lambda$ all perform well; specifically, this procedure with $\lambda = 2\alpha$ performs the most powerful among all procedures, whereas the DAG BH performs the worst. This shows that for the DAG GELS procedure, $\lambda= 2\alpha$ is a good choice, but for other values of $\lambda$, it does not weaken its power performance so much. In terms of interpretation, we need to note that the rejections by both the DAG GELS procedure and the DAG BH procedure preserve the underlying GO structure and perhaps have a more natural interpretation.

\begin{table}
    \caption{The number of rejections out of 10,362 hypotheses by the DAG GELS procedure with $\lambda=\alpha, 2\alpha, 4\alpha$ and $10\alpha$, the DAG BH, and the BH procedure at various significance levels for the leukemia microarray data set of Golub et al. (1999).}

    \vspace{-5pt}
    \begin{center}

    \begin{tabular}{ccccccc}
        \hline
        \textbf{$\alpha$} & \textbf{DAG GELS} & \textbf{DAG GELS} & \textbf{DAG GELS} & \textbf{DAG GELS} &\textbf{DAG BH} & \textbf{BH}\\
                          & \textbf{$\lambda=\alpha$} & \textbf{$\lambda=2\alpha$} & \textbf{$\lambda=4\alpha$} & \textbf{$\lambda=10\alpha$} & & \\ \noalign{\hrule height 2pt}
          0.0001  & 2226  & 2226  & 2226  & 2226  & 1882  & 2170  \\ \hline
          0.001  & 3226  & 3226  & 3226  & 3226  & 2746  & 3086  \\ \hline
          0.01  & 4219  & 4368  & 4349  & 4322  & 3947  & 4245  \\ \hline
          0.025  & 4779  & 5066  & 5027  & 4935  & 4594  & 4898  \\ \hline
          0.05  & 5292  & 5705  & 5632  & 5461  & 5210  & 5542  \\ \hline
          0.1  & 5856  & 6384  & 6243  & 5973  & 6017  & 6258  \\ \hline
    \end{tabular}
    \label{TABLE_DAG_REJECTIONS}
    \end{center}
\end{table}

\section{Conclusion}

In this paper, we have provided a new general approach for developing procedures controlling large scale error rates such as the FDR. This new approach can be used to prove FDR control of existing procedures such as the adaptive BH procedure and can also be used to develop new procedures controlling other error rates such as the generalized FDR, as illustrated in Examples \ref{EXMP_ADAPT_FDR} and \ref{EXMP_GENERALIZED_FDR}.

As a consequence of this framework, under the assumption of independence,
we have developed a new FDR controlling procedure for testing hypotheses along a DAG, which was termed as the DAG GELS procedure. In the simulation study, this DAG GELS procedure was shown to be more powerful than the BH procedure by accounting for the underlying DAG structure. Moreover, the set of rejected hypotheses by the DAG GELS procedure preserves the graphical structure which may enhance interpretation. Finally, because the DAG structure is a very general structure, the DAG GELS procedure also has applications towards the testing of hypotheses that have a hierarchical or fixed sequence structure.

We need to note that the
general approach was developed under the assumption of positive dependence or independence. An interesting future work is to generalize this approach to the case of arbitrary dependence and based on the generalized approach, to develop new procedures controlling the FDR or other error rates under arbitrary dependence for various settings of multiple testing.


\section{Proofs}

\subsection{Proof of Lemma \ref{LEMMA_POS_INDPENDENCE}}

Suppose $L$ takes the form $L(\alpha) = \ind{\bigcap_{i=1}^{m}P_i \le t_i(\alpha)}$ for non-decreasing $t_i, i = 1, \dots, m$. Given a coordinatewise non-decreasing function of the $p$-values $\psi$, define the functions
\begin{eqnarray}
&& g_i(\alpha, p_1, \dots, p_{i-1}) \nonumber \\
&=& \expv{\psi(P_1, \dots, P_m) \cond \bigcap_{j=i}^{m}P_j \le t_j(\alpha), P_1 = p_1, \dots, P_{i-1} = p_{i-1}} \nonumber
\end{eqnarray}
for $i = 1, \dots, m$. It is easy to see that when the $p$-values are independent, $g_i(\alpha, p_1, \dots, p_{i-1})$ is a coordinatewise non-decreasing function of $p_1, \dots, p_{i-1}$. In the following, we will show that it is also a non-decreasing function of $\alpha$ when $p_j \le t_j(\alpha)$ for $j = 1, \dots, i-1$.

Let $0 < \alpha \le \alpha'$. For $p_j \le t_j(\alpha), j = 1, \dots, m-1$, we have that
\begin{eqnarray}
    && g_m(\alpha, p_1, \dots, p_{m-1}) \nonumber \\
&=& \expv{\psi(P_1, \dots, P_m) \cond P_m \le t_m(\alpha), P_1 = p_1, \dots, P_{m-1} = p_{m-1}} \nonumber \\
&\le& \expv{\psi(P_1, \dots, P_m) \cond P_m \le t_m(\alpha'), P_1 = p_1, \dots, P_{m-1} = p_{m-1}} \nonumber \\
&=& g_m(\alpha', p_1, \dots, p_{m-1}). \nonumber
\end{eqnarray}
The inequality follows by independence of the $p$-values and by the fact that $\psi$ is a non-decreasing function of $P_m$ and $t_m(\cdot)$ is a non-decreasing function. Thus, $g_m$ is a non-decreasing function of $\alpha$. By induction, assume $g_{i+1}$ is a non-decreasing function of $\alpha$ for $p_j \le t_j(\alpha), j = 1, \dots, i$. Then,
\begin{eqnarray}
    && g_i(\alpha, p_1, \dots, p_{i-1})
=    \expv{g_{i+1}(\alpha, p_1, \dots, p_{i-1}, P_i) \cond P_i \le t_i(\alpha)} \nonumber \\
&\le& \expv{g_{i+1}(\alpha', p_1, \dots, p_{i-1}, P_i) \cond P_i \le t_i(\alpha)} \nonumber \\
&\le& \expv{g_{i+1}(\alpha', p_1, \dots, p_{i-1}, P_i) \cond P_i \le t_i(\alpha')} \nonumber \\
&=& g_i(\alpha', p_1, \dots, p_{i-1}). \nonumber
\end{eqnarray}
The first inequality follows by induction. The second inequality follows by independence of the $p$-values and by the fact that $g_{i+1}$ is a non-decreasing function of $P_i$ and $t_i(\cdot)$ is a non-decreasing function. Therefore, $g_1(\alpha)$ is non-decreasing in $\alpha$ and then the desired result follows by the fact that $g_1(\alpha) = \expv{\psi(P_1, \dots, P_m) \cond L(\alpha) > 0}$. ~\qed

\subsection{Proof of Theorem \ref{THM_COMPOSITE}}

By condition C1, we suppose that $L_1, \dots, L_s$ are the binary error rates such that $\sum_{i=1}^s L_i \ge L$, where $s$ is a given positive integer. Based on Definition \ref{GELS procedure}, we have that the base procedure is applied at level $\alpha/W(R)$, where $R$ is the number of rejections. Thus, for each $i = 1, \dots, s$,
\begin{eqnarray}\label{Li_ineq}
    && \expv{W(R) L_i (\alpha/W(R))} = \sum_{r=1}^{m}W(r)L_i(\alpha/W(r))\pr{R = r, L_i(\alpha/W(r)) > 0} \nonumber \\
&\le& \sum_{r=1}^{m}c_i\alpha\pr{R = r \cond L_i(\alpha/W(r)) > 0} \nonumber \\
&=& \sum_{r=1}^{m}c_i\alpha\pr{R \ge r \cond L_i(\alpha/W(r)) > 0} - \sum_{r=1}^{m-1}c_i\alpha\pr{R \ge r+1 \cond L_i(\alpha/W(r)) > 0} \nonumber \\
&\le& c_i\alpha + \sum_{r=1}^{m-1}c_i\alpha\left[\pr{R \ge r+1 \cond L_i(\alpha/W(r+1)) > 0} \right. \nonumber \\
&& \qquad \qquad \qquad \qquad \left. - \pr{R \ge r+1 \cond L_i(\alpha/W(r)) > 0}\right] \nonumber \\
&\le& c_i\alpha.
\end{eqnarray}
The first inequality follows by condition C2 and by the fact that $L_i$ is a binary error rate. The last inequality follows by condition C3 and by the fact that $W(\cdot)$ is a non-increasing function.

Therefore, by condition C1 and (\ref{Li_ineq}), we have
\begin{eqnarray}
E\left[W(R) L(\alpha/W(R))\right] \le \sum_{i=1}^{s}E\left[W(R) L_i(\alpha/W(R))\right]
\le \sum_{i=1}^{s}c_i\alpha \le \alpha, \nonumber
\end{eqnarray}
the desired result. ~\qed

\subsection{Proof of (\ref{EQN_KFDR})}
Let $I_0$ denote the index set of true null hypotheses and $V^{(-\mathcal{J})}$ the number of false rejections excluding all $H_j$ for $j \in \mathcal{J}$. Then
\begin{eqnarray}
L &=& V\ind{V \ge k} =   \sum_{i \in I_0}\ind{P_i \le t, V \ge k} \nonumber \\
&=&    \sum_{\mathcal{J} \in \mathcal{C}_1^0}\ind{\max_{j \in \mathcal{J}}P_j \le t, V^{(-\mathcal{J})} \ge k-1} \nonumber \\
&\le&  \sum_{\mathcal{J} \in \mathcal{C}_1^0}\frac{V^{(-\mathcal{J})}}{k-1}\ind{\max_{j \in \mathcal{J}}P_j \le t, V^{(-\mathcal{J})} \ge k-1} \nonumber \\
&=&  \sum_{\mathcal{J} \in \mathcal{C}_1^0}\sum_{i \in I_0\setminus \mathcal{J}}\frac{1}{k-1}\ind{\max_{j \in \mathcal{J}}P_j \le t, P_i \le t, V^{(-\mathcal{J})} \ge k-1} \nonumber \\
&=&  \sum_{\mathcal{J} \in \mathcal{C}_2^0}\frac{1}{k-1}\ind{\max_{j \in \mathcal{J}}P_j \le t, V^{(-\mathcal{J})} \ge k-2} \nonumber \\
&\le&  \cdots \quad \le \nonumber \\
&\le& \sum_{\mathcal{J} \in \mathcal{C}_k^0}\frac{1}{(k-1) !}\ind{\max_{j \in \mathcal{J}}P_j \le t} \nonumber \\
&=&    \sum_{\mathcal{J} \in \mathcal{C}_k^0}L_\mathcal{J}, \nonumber
\end{eqnarray}
the first inequality of (\ref{EQN_KFDR}). Here, the first inequality follows by the fact that the event $\{V^{(-\mathcal{J})} \ge k-1\}$ implies $V^{(-\mathcal{J})} / (k-1) \ge 1$.

To show the second inequality of (\ref{EQN_KFDR}), note that for any $\mathcal{J} \in \mathcal{C}_k^0$,
\begin{eqnarray}
&& \expv{L_\mathcal{J}} = \frac{1}{(k-1) !}\pr{\max_{j \in \mathcal{J}}P_j \le t} \nonumber \\
&=&   \frac{1}{(k-1) !}\prod_{j \in \mathcal{J}}\pr{P_j \le t} \nonumber \\
&\le& \frac{1}{(k-1) !}t^k =   c_\mathcal{J} \alpha \nonumber
\end{eqnarray}
and $|\mathcal{C}_k^0| = \binom{m_0}{k}$, thus $\sum_{\mathcal{J} \in \mathcal{C}_k^0}c_\mathcal{J} = \binom{m_0}{k}/\binom{m}{k} \le 1$, the desired result. ~\qed


\subsection{Proof of Lemma \ref{LEM2}}

\emph{Proof of (\ref{property1}).} For $H_j \in \mathcal{F}_1$,  we have $\sum_{H_k \in \mathcal{F}_1}s_{k, j} = \sum_{H_k \in \mathcal{F}_1}\ind{k = j} = 1$ and (\ref{property1}) holds. By induction, assume (\ref{property1}) holds for all $j$ such that $H_j \in \mathcal{F}_{t}$, then for $H_j \in \mathcal{F}_{t+1}$, we have
\[
   \sum_{H_k \in \mathcal{F}_1}s_{k, j}
=  \sum_{H_k \in \mathcal{F}_1}\sum_{H_{k'} \in \mathcal{T}_j}\frac{s_{k, {k'}}}{|\mathcal{T}_j|}
=  \sum_{H_{k'} \in \mathcal{T}_j}\sum_{H_k \in \mathcal{F}_1}\frac{s_{k, {k'}}}{|\mathcal{T}_j|}
=  \sum_{H_{k'} \in \mathcal{T}_j}\frac{1}{|\mathcal{T}_j|}
=  1.
\]
The first equality follows by (\ref{EQN_S_DAG}) and the third follows by induction, since $H_{k'} \in \mathcal{T}_j$ and $H_j \in \mathcal{F}_{t+1}$, then $H_{k'} \in \bigcup_{u=1}^t\mathcal{F}_u$. Thus, (\ref{property1}) holds for $j = 1, \ldots, m$.

\vspace{10pt}
\noindent \emph{Proof of (\ref{property2}).}
Note that (\ref{property2}) holds trivially if $H_j \notin  \mathcal{M}_i$.
For $H_j \in  \mathcal{M}_i$,  suppose $H_i \in \mathcal{F}_{t}$ for some
$t = 1, \ldots, m$. Since $i \neq j$, thus $H_j \in \mathcal{F}_{u}$ with some $u > t$. For $H_j \in \mathcal{F}_{t+1}$, we have $H_i \in \mathcal{T}_j$, which follows by
$H_j \in  \mathcal{M}_i$ and $H_i \in \mathcal{F}_{t}$. Thus, (\ref{property2}) holds since it can be seen that (\ref{EQN_S_DAG}) gives $s_{i, j} = 1/|\mathcal{T}_j|$, and $s_{k, j} = I\{k = j\}$ for each $k$ such that $H_i \in \mathcal{T}_k$. By induction, assume (\ref{property2}) holds for all $j$ such that $H_j \in \bigcup_{\ell = t+1}^{u-1}\mathcal{F}_{\ell}$ for some $u > t +1$. Then, for $H_j \in \mathcal{F}_{u}$, we have
\begin{eqnarray}
s_{i, j} &=& \sum_{H_k \in \mathcal{T}_j}\frac{s_{i, k}}{|\mathcal{T}_j|} = \sum_{\substack{H_k \in \mathcal{T}_j}}\sum_{\ell: H_i \in \mathcal{T}_\ell}\frac{s_{\ell, k}}{|\mathcal{T}_j||\mathcal{T}_\ell|} \nonumber \\
&=& \sum_{\substack{\ell : H_i \in \mathcal{T}_\ell }}\sum_{H_k \in \mathcal{T}_j}\frac{s_{\ell, k}}{|\mathcal{T}_j||\mathcal{T}_\ell|} = \sum_{\substack{\ell: H_i \in \mathcal{T}_\ell}}\frac{s_{\ell, j}}{|\mathcal{T}_\ell|}. \nonumber
\end{eqnarray}
The first equality follows by (\ref{EQN_S_DAG}).
The second equality follows by induction where it should be noted that $H_k \in \mathcal{T}_j$ and $H_j \in \mathcal{F}_{u}$ imply $H_k \in \bigcup_{v=1}^{u-1}\mathcal{F}_v$ so that induction is valid for
$H_k \in \bigcup_{v=t+1}^{u-1}\mathcal{F}_v$. For $H_k \in \bigcup_{v=1}^{t}\mathcal{F}_v$, if $H_i \in \mathcal{T}_j$, then the second equality follows by the same argument as in the case of
$H_j \in \mathcal{F}_{t+1}$; if $H_i \notin \mathcal{T}_j$, then
$s_{i, k} = 0$ and $s_{\ell, k} = 0$
for all $\ell$ such that $H_i \in \mathcal{T}_\ell$, thus the second equality holds trivially. The fourth equality also follows by (\ref{EQN_S_DAG}). Thus, (\ref{property2}) holds for $i, j = 1, \ldots, m$.  ~\qed

\subsection{Proof of Theorem \ref{THM_PFER_TESTING_PROC}}

Theorem \ref{THM_PFER_TESTING_PROC} can be regarded as a corollary of Theorem \ref{THM_DAG_TESTING_PROC} with $W(R) = 1$. This completes the proof of
Theorem \ref{THM_PFER_TESTING_PROC} upon the following proof of Theorem \ref{THM_DAG_TESTING_PROC}. ~\qed

\subsection{Proof of Theorem \ref{THM_DAG_TESTING_PROC}}

To show that $\expv{W(R)L}$ is controlled, by Theorem \ref{THM_COMPOSITE}, it is enough to show that the DAG testing procedure with the critical constants $\alpha_i, ~i = 1, \ldots, m$ satisfies conditions C1-C3 under independence. Note that $V$ can be split into $m$ error rates $L_i = \ind{H_i \text{ is falsely rejected}}$, $i = 1, \dots, m$. Since $H_i$ cannot be tested unless all of its ancestors are rejected, we have
\[
L_i = \ind{\bigcap_{j \in \mathcal{D}_i}P_j \le \alpha_j, H_i \text{ is true}}, i = 1, \dots, m.
\]
Condition C1 is satisfied since $L_i$ is a binary error rate and is a non-increasing function of the $p$-values. By Lemma \ref{LEMMA_POS_INDPENDENCE}, we also have that condition C3 is satisfied under independence. Finally, we will show that condition C2 is satisfied. For each $i = 1, \dots, m$, if $H_i$ is false, then let $c_i = 0$ and if $H_i$ is true, then let
\[
c_i = \dfrac{\ell_i}{\ell(1+\lambda\ell_i)^{\ind{|\mathcal{M}_i| > 1}}}\prod_{\substack{j \in \mathcal{D}_i \\ j \neq i}}\left(\frac{\lambda \ell_j}{1 + \lambda\ell_j}\right)^{\ind{H_j \text{ is true}}}.
\]

To show that condition C2 is satisfied, we must show the following two results:
\begin{enumerate}
\item[(a).] $\expv{L_i(\alpha)} \le c_i\alpha$ for each $i = 1, \dots, m$.
\item[(b).] $\sum_{i=1}^{m}c_i \le 1.$
\end{enumerate}

\vspace{10pt}
\emph{Proof of part (a).} If $H_i$ is false, then $L_i = 0$ and clearly part (a) holds. If $H_i$ is true, then
\begin{eqnarray}
&& \expv{L_i} = \pr{\bigcap_{j \in \mathcal{D}_i}P_j \le \alpha_j} \nonumber \\
&\le&  \pr{P_i \le \dfrac{\ell_i \alpha}{\ell(1+\lambda\ell_i)^{\ind{|\mathcal{M}_i| > 1}}}, \bigcap_{\substack{j \in \mathcal{D}_i \\ j \neq i}}P_j \le \dfrac{\lambda \ell_j}{1 + \lambda\ell_j}} \nonumber \\
&\le&  \dfrac{\ell_i \alpha}{\ell(1+\lambda \ell_i)^{\ind{|\mathcal{M}_i| > 1}}}\prod_{\substack{j \in \mathcal{D}_i \\ j \neq i}}\left(\frac{\lambda \ell_j}{1 + \lambda \ell_j}\right)^{\ind{H_j \text{ is true}}} = c_i \alpha. \nonumber
\end{eqnarray}
The first inequality follows from the fact that $\ell_i = 1$ if $H_i$ is a leaf hypothesis. The second inequality follows by independence of the $p$-values.

\vspace{10pt}
\emph{Proof of part (b).} Part (b) follows by Lemma \ref{LEM2} (i) and the following inequality:
\begin{equation}
\sum_{H_j \in \mathcal{M}_i}s_{i, j}c_j \le \frac{\ell_i}{\ell}, ~H_i \in \mathcal{F}_1. \label{EQN_DAG_2}
\end{equation}
Then, we have
\begin{eqnarray}
&& \sum_{j = 1}^{m}c_j = \sum_{j = 1}^{m}\sum_{H_i \in \mathcal{F}_1}s_{i, j}c_j =  \sum_{H_i \in \mathcal{F}_1}\sum_{H_j \in \mathcal{M}_i}s_{i, j}c_j \nonumber \\
&\le& \sum_{H_i \in \mathcal{F}_1}\frac{\ell_i}{\ell}  =  \sum_{H_i \in \mathcal{F}_1}\sum_{H_j \text{ is a leaf}}\frac{s_{i, j}}{\ell} = \sum_{H_j \text{ is a leaf}}\frac{1}{\ell} = 1. \nonumber
\end{eqnarray}
The first equality follows by Lemma \ref{LEM2} (i) and the second follows by the fact that $s_{i, j} = 0$ if $H_j \notin \mathcal{M}_i$. The inequality follows by (\ref{EQN_DAG_2}). The third equality follows by the definition of $\ell_i$ and the fourth follows by Lemma \ref{LEM2} (i) again.

\vspace{5pt}
\emph{Proof of (\ref{EQN_DAG_2}).} To show that (\ref{EQN_DAG_2}) holds, we will show that the following inequality holds.
\begin{equation}
\sum_{H_j \in \mathcal{M}_i}s_{i, j}c_j \le \frac{\ell_i}{\ell} \prod_{\substack{j \in \mathcal{D}_i \\ j \neq i}}\left(\frac{\ell_j \lambda}{1 + \ell_j \lambda}\right)^{\ind{H_j \text{ is true}}}, ~i =1 , \dots, m. \label{EQN_DAG_7}
\end{equation}
It should be noted that in the special case when $H_i \in \mathcal{F}_1$, (\ref{EQN_DAG_7}) reduces to (\ref{EQN_DAG_2}).

Suppose that $H_i$ is a leaf hypothesis so that $\mathcal{M}_i = \{H_i\}, |\mathcal{M}_i| = 1$ and $\ell_i = 1$. If $H_i$ is false, (\ref{EQN_DAG_7}) holds trivially since $c_i = 0$. If $H_i$ is true, then
\begin{equation}
    \sum_{H_j \in \mathcal{M}_i}s_{i, j}c_j
=   c_i
=   \frac{\ell_i}{\ell} \prod_{\substack{j \in \mathcal{D}_i \\ j \neq i}}\left(\frac{\ell_j \lambda}{1 + \ell_j \lambda}\right)^{\ind{H_j \text{ is true}}}. \label{EQN_DAG_6}
\end{equation}
Thus, (\ref{EQN_DAG_7}) holds for all leaf hypotheses $H_i$.

Suppose that $H_i$ is not a leaf hypothesis so that $|\mathcal{M}_i| > 1$. By induction, assume that (\ref{EQN_DAG_7}) holds for all descendant hypotheses $H_k$ of $H_i$, i.e., $H_k \in \mathcal{M}_i\setminus\{H_i\}$. Then,
\begin{eqnarray}
&&    \sum_{H_j \in \mathcal{M}_i}s_{i, j}c_j
 ~=~   c_i + \sum_{\substack{H_j \in \mathcal{M}_i \\ i \neq j}}s_{i, j}c_j
 ~=~   c_i + \sum_{H_j \in \mathcal{M}_i}\sum_{k : H_i \in \mathcal{T}_k}\frac{s_{k, j}}{|\mathcal{T}_k|}c_j \nonumber \\
&=&   c_i + \sum_{k : H_i \in \mathcal{T}_k}\sum_{H_j \in \mathcal{M}_k}\frac{s_{k, j}}{|\mathcal{T}_k|}c_j
~\le~ c_i + \sum_{k : H_i \in \mathcal{T}_k}\frac{1}{|\mathcal{T}_k|}\frac{\ell_k}{\ell} \prod_{\substack{j \in \mathcal{D}_k \\ j \neq k}}\left(\frac{\ell_j \lambda}{1 + \ell_j \lambda}\right)^{\ind{H_j \text{ is true}}} \nonumber  \\
&\le& c_i + \sum_{k : H_i \in \mathcal{T}_k}\frac{1}{|\mathcal{T}_k|}\frac{\ell_k}{\ell} \prod_{\substack{j \in \mathcal{D}_i}}\left(\frac{\ell_j \lambda}{1 + \ell_j \lambda}\right)^{\ind{H_j \text{ is true}}}. \label{EQN_DAG_3}
\end{eqnarray}
The second equality follows by Lemma \ref{LEM2} (ii) and the fact that $s_{k, i} = 0$ for all $k$ such that $H_i \in \mathcal{T}_k$. The third equality follows from the fact that $s_{k, j} = 0$ if $H_j \notin \mathcal{M}_k$. The first inequality follows by induction and the second by the fact that $\mathcal{D}_i \subseteq \mathcal{D}_k \setminus \{H_k\}$ for $H_i \in \mathcal{T}_k$.

Note that
\begin{eqnarray} \label{EQN_DAG_32}
&& \sum_{k:H_i \in \mathcal{T}_k}\dfrac{\ell_k}{|\mathcal{T}_k|} = \sum_{k:H_i \in \mathcal{T}_k}\sum_{H_j \text{ is a leaf}}\dfrac{s_{k, j}}{|\mathcal{T}_k|} \nonumber \\
&=& \sum_{H_j \text{ is a leaf}}\sum_{k:H_i \in \mathcal{T}_k}\dfrac{s_{k, j}}{|\mathcal{T}_k|} = \sum_{H_j \text{ is a leaf}}s_{i, j} = \ell_i.
\end{eqnarray}
The first and last equalities follow from the definition of $\ell_i$ and the third follows by
Lemma \ref{LEM2} (ii). Then, by (\ref{EQN_DAG_32}), the right-hand side of (\ref{EQN_DAG_3}) can be simplified to
\begin{equation}
c_i + \frac{\ell_i}{\ell} \prod_{\substack{j \in \mathcal{D}_i}}\left(\frac{\ell_j \lambda}{1 + \ell_j \lambda}\right)^{\ind{H_j \text{ is true}}}.  \label{EQN_DAG_4}
\end{equation}

If $H_i$ is false, then $c_i = 0$ and by (\ref{EQN_DAG_4}), it is easy to see that (\ref{EQN_DAG_7}) holds. If $H_i$ is true, then by the definition of $c_i$ and the fact that $|\mathcal{M}_i| > 1$, (\ref{EQN_DAG_4}) can be expressed as
\begin{eqnarray}
&& \dfrac{\ell_i}{\ell(1+\ell_i\lambda)}\prod_{\substack{j \in \mathcal{D}_i \\ j \neq i}}\left(\frac{\ell_j\lambda}{1 + \ell_j\lambda}\right)^{\ind{H_j \text{ is true}}} + \frac{\ell_i}{\ell} \prod_{\substack{j \in \mathcal{D}_i}}\left(\frac{\ell_j \lambda}{1 + \ell_j \lambda}\right)^{\ind{H_j \text{ is true}}} \nonumber \\
&=& \frac{1}{\lambda \ell}\prod_{\substack{j \in \mathcal{D}_i}}\left(\frac{\ell_j \lambda}{1 + \ell_j \lambda}\right)^{\ind{H_j \text{ is true}}} + \frac{\ell_i}{\ell} \prod_{\substack{j \in \mathcal{D}_i}}\left(\frac{\ell_j \lambda}{1 + \ell_j \lambda}\right)^{\ind{H_j \text{ is true}}} \nonumber \\
&=&  \left(\frac{1}{\lambda \ell} + \frac{\ell_i}{\ell} \right)\prod_{j \in \mathcal{D}_i}\left(\frac{\ell_j \lambda}{1 + \ell_j \lambda}\right)^{\ind{H_j \text{ is true}}} \nonumber \\
&=&  \frac{\ell_i}{\ell}\prod_{\substack{j \in \mathcal{D}_i \\ j \neq i}}\left(\frac{\ell_j \lambda}{1 + \ell_j \lambda}\right)^{\ind{H_j \text{ is true}}}. \nonumber
\end{eqnarray}
This completes the proof of (\ref{EQN_DAG_7}) and (\ref{EQN_DAG_2}) then follows. ~~\qed

\subsection{Proof of Proposition \ref{PROP_DAG_BH}}

Let $L_i(\alpha) = \ind{P_i \le \alpha/m, H_i \text{ is true}}$, then
\begin{eqnarray}
V &=& \sum_{i=1}^{m}\ind{H_i \text{ is falsely rejected}} \nonumber \\
&\le& \sum_{i=1}^{m}\ind{P_i \le \alpha/m, H_i \text{ is true}}, \nonumber
\end{eqnarray}
and thus condition C1 is satisfied. Let $c_i = 1/m$ and note that $\expv{L_i(\alpha)} \le \alpha/m = c_i\alpha$, thus condition C2 is satisfied. Finally, condition C3 follows directly from Assumption \ref{ASM_POS_DEPENDENCE}.
~\qed




\end{document}